\documentclass[twocolumn,aps,showpacs]{revtex4}
\usepackage{graphicx}

\begin{document}

%
%

\title{Exact wave functions for the edge state of a disk-shaped two dimensional topological insulator}

\author{M. Pang and X. G. Wu}

\affiliation{SKLSM, Institute of Semiconductors, Chinese Academy
of Sciences, Beijing 100083, China}

\begin{abstract}

We report the exact wave functions for the eigen state of a disk-shaped two
dimensional topological insulator.  The property of the edge state whose energy
lies inside the bulk gap is studied.  It is found that the edge state energy is
affected by the radius of the disk.  For a fixed angular momentum index, there
is a critical disk radius below which there exists no edge state.  The value of
this critical radius increases as the angular momentum index increases.
In the limit of large disk radius, the energy of the edge state approaches a
limiting value determined by the system parameters and independent of the
angular momentum index.  The derivation from this limiting value is inversely
proportional to the radius with a coefficient proportional to the angular
momentum index.  In the general case, the energy differences between two edge states
with adjacent angular momentum indexes are not equal.  The exact and analytical
wave functions also facilitates the investigation of electronic state in
other structures of the two dimensional topological insulator.

\end{abstract}


\maketitle

%
%

%
%

Topological insulator has attracted considerable attentions in recent years
\cite{Zhang08Phys,Hasan10RevModPhys,Moore10Nature}.  In two dimensions, the
topological insulator is described by an effective model \cite{Zhang06}.
Many theoretical investigations into the exotic properties of two dimensional
topological insulator (2DTI) are based on this model
\cite{Niu08,Liu08PRL,Li09PRL,Schmidt09PRB,Chang11PRL}.

Despite those great progresses achieved, exact wave function for the 2DTI is
rarely reported so far.  When a 2DTI is cut into a ribbon like structure, exact wave
functions were obtained \cite{Niu08}.  It is found that due to the finite
width of ribbon, there is a gap in the energy spectrum which decreases
exponentially as the width of the ribbon increases \cite{Niu08}.

In the present paper, we report exact wave functions for the 2DTI with a
circular geometry.  This allows us to probe some exact properties of 2DTI.
In particular, we focus on the edge state of a disk shaped 2DTI with the
open boundary condition that both components of wave functions vanish at
the disk edge.  The electronic state in other geometric structure will also
be briefly discussed.

%
%

The 2DTI is described by the following well-known Hamiltonian \cite{Zhang06}
\begin{equation}\label{effmodel}
   \left(
   \begin{array}{*{20}c}
   {h( {\bf k} )}  &  {0}  \\
   {0}  &  { h^*(-{\bf k}) }  \\
   \end{array}
   \right) ,
\end{equation}
where ${\bf k}=(k_x,k_y)$.  $h({\bf k})=\epsilon({\bf k})+\sum_\alpha
d_\alpha({\bf k})\sigma_\alpha$
with $\sigma_\alpha$ the Pauli matrix.  $\epsilon({\bf k})=C-D(k_x^2+k_y^2)$,
$d_1=Ak_x$, $d_2=Ak_y$, and $d_3=M-B(k_x^2+k_y^2)$.  $A$, $B$, $C$, $D$,
and $M$ are parameters determined by the structure of the quantum
well \cite{Zhang06}.  Operators $k_x$ and $k_y$ represent differential
operators $-i\partial_x$ and $-i\partial_y$ respectively.  The parameter $C$
gives the zero point of energy and we can safely set it to zero for
simplicity in this paper.
The spin-up block $h( {\bf k} )$ and spin-down block $h^*(-{\bf k})$ in the
Hamiltonian are decoupled and they can be solved separately \cite{Zhang06}.
Since the the system under consideration has a circular geometry, we will
adopt the polar coordinate system.  It can be shown that the radial and
angular part of the wave function can be separated.  This simplifies the
problem and one only needs to solve the radial wave function.

For the spin-up block $h( {\bf k} )$, the wave function can be written as
\begin{equation}
  \Psi_m(r,\theta)=
    \left(\begin{array}{*{20}c}
    {e^{im\theta} \phi_1(r)}    \\
    {e^{i(m+1)\theta} \phi_2(r)}    \\
    \end{array}\right) ,
\end{equation}
where $m=0, \pm 1, \pm2, \pm 3, ...$ is the angular momentum index.
The differential equations for the radial part can be written as
\begin{equation}\label{ham}
    \left( \begin{array}{*{20}c}
    {O_m+X_1} & {X_3P_{m+1} } \\
    {X_4P_{-m} } & {O_{m+1}+X_2} \\
    \end{array} \right)
    \left( \begin{array}{*{20}c}
    {\phi_1(r)} \\
    {\phi_2(r)} \\
    \end{array} \right) = 0 ,
\end{equation}
with $O_m={d^2/d^2r}+{d/rdr}-{m^2/r^2}$, $P_m=d/dr+m/r$,
$X_1=(-E+V(r)+M)/(B+D)$, $X_2=(E-V(r)+M)/(B-D)$, $X_3=-iA/(B+D)$, and
$X_4=iA/(B-D)$.  $E$ is the eigen energy of the system and $V(r)$ the
externally applied potential that depends only on the radial coordinate.

In this paper, we consider $V(r)$ taking different but constant
values in the different regions of $r$.  In this case, the exact
wave function can be written as
\begin{equation}\label{solt}
    \Phi_m(\lambda r)=
        \left(
        \begin{array}{*{20}c}
        {\phi_1(r)} \\
        {\phi_2(r)} \\
        \end{array}
        \right) =
    \left(\begin{array}{*{20}c}
    {C_1Z_m(\lambda r)} \\
    {C_2Z_{m+1}(\lambda r)} \\
    \end{array}\right) ,
\end{equation}
with $Z_m(\lambda r)$ the Bessel functions $J_m(\lambda r)$ or $Y_m(\lambda r)$.
$\lambda$, $C_1$, and $C_2$ are solution of a secular equation
\begin{equation}
    \left(
    \begin{array}{*{20}c}
    X_1 - \lambda^2 & X_3 \lambda \\
    -X_4 \lambda & X_2 - \lambda^2 \\
    \end{array}
    \right)
    \left(
    \begin{array}{*{20}c}
    C_1(\lambda) \\
    C_2(\lambda) \\
    \end{array}
    \right) = 0 .
\end{equation}
The equation gives four roots of $\lambda$: $\pm \lambda_1$ and $\pm \lambda_2$,
as
\begin{equation}\label{lam}
\begin{array}{c}
  \lambda_1 = [ ( F + {\sqrt{F^2 - 4 X_1 X_2} } )/2 ]^{1/2} ,
  \\
  \lambda_2 = [ ( F - {\sqrt{F^2 - 4 X_1 X_2} } )/2 ]^{1/2} ,
\end{array}
\end{equation}
with $F=X_1+X_2-X_3X_4$.  Note that $\lambda$ is a function of potential $V$
and energy $E$.  $C_1$ and $C_2$ are only determined up to factor.

The value of $\lambda$ obtained in Eq.(\ref{lam}) can become imaginary.
In that case, it is more convenient to write the exact wave function as
\begin{equation}
    \Phi_m(\xi r)=
        \left(
        \begin{array}{*{20}c}
        {\phi_1(r)} \\
        {\phi_2(r)} \\
        \end{array}
        \right) =
    \left(\begin{array}{*{20}c}
    {C_3 Z_m(\xi r)} \\
    {C_4 Z_{m+1}(\xi r)} \\
    \end{array}\right) ,
\end{equation}
with $Z_m(\xi r)$ the Bessel functions $I_m(\xi r)$ or $(-1)^mK_m(\xi r)$.
$\xi$, $C_3$, and $C_4$ are solution of a secular equation
\begin{equation}
    \left(
    \begin{array}{*{20}c}
    X_1 + \xi^2 & X_3 \xi \\
    X_4 \xi & X_2 + \xi^2 \\
    \end{array}
    \right)
    \left(
    \begin{array}{*{20}c}
    C_3(\xi) \\
    C_4(\xi) \\
    \end{array}
    \right) = 0 .
\end{equation}
The equation gives four roots of $\xi$: $\pm \xi_1$ and $\pm \xi_2$,
as
\begin{equation}\label{xi}
\begin{array}{c}
  \xi_1 = [ ( - F + {\sqrt{F^2 - 4 X_1 X_2} } )/2 ]^{1/2} ,
  \\
  \xi_2 = [ ( - F - {\sqrt{F^2 - 4 X_1 X_2} } )/2 ]^{1/2} ,
\end{array}
\end{equation}
with $F=X_1+X_2-X_3X_4$ the same as given before.  The $E$ dependence
of $\xi$ in Eq.(\ref{xi}) depends also on other model parameters.
With the parameters given in \cite{Zhang06}, one can readily verify that
when $|E-V|<|M|$, both $\xi_1$ and $\xi_2$ are real and non-zero.
When $|E-V|=|M|$, $\xi_1$ is non-zero and real, and $\xi_2=0$.
When $|E-V|>|M|$, both $\xi_1$ and $\xi_2$ are non-zero, $\xi_1$
remains real, and $\xi_2$ is purely imaginary.

It can be shown that $\Phi_m(\lambda r)$ and $\Phi_m(-\lambda r)$
(or $\Phi_m(\xi r)$ and $\Phi_m(-\xi r)$) are not linearly independent solutions.
Therefore we have linearly independent solutions $\Phi^J_m(\lambda r)$
and $\Phi^Y_m(\lambda r)$, or $\Phi^I_m(\xi r)$ and $\Phi^K_m(\xi r)$.
The superscripts $J$, $Y$, $I$, and $K$ denotes the kind of the Bessel
functions involved.  The desired wave function can be constructed as a linear
combination of the linearly independent solutions.  In the case of $\lambda=0$
or $\xi=0$, a careful treatment of $\lambda\to 0$ or $\xi\to 0$ limit
is required in order to obtain linearly independent solutions.

For the spin-down block $h^*(-{\bf k})$, the exact wave function can be
obtained in the same way.  A careful examination shows that the energy of
spin-up state with angular momentum index $-|m|$, denoted as $E_{-|m|,\uparrow}$,
exactly equals to $E_{|m|-1,\downarrow}$ the energy of spin-down state with
angular momentum index $|m|-1$.  One has $E_{-|m|,\uparrow}=E_{|m|-1,\downarrow}$
and $E_{-|m|,\downarrow}=E_{|m|-1,\uparrow}$.

%
%

Next, we use the exact wave functions obtained above to construct the edge
state for a disk shaped 2DTI.  In this quantum disk system, one has $V(r)=0$
for $r<R$ with $R$ the radius of the disk.  The boundary condition is that
both components of the wave functions must vanish at $r=R$.  When the energy
falls inside the bulk gap $E^2<M^2$, Eq.(\ref{xi}) gives to two
real $\xi_1$ and $\xi_2$ with model parameters given in Ref.\cite{Zhang06}.
The wave functions that involve the Bessel function $K_m(\xi r)$ are
divergent at $r=0$ and must be discarded.  Therefore, we construct the
desired wave function from $\Phi^I_m(\xi_1 r)$ and $\Phi^I_m(\xi_2 r)$
by requiring $d_1\Phi^I_m(\xi_1 R)+d_2\Phi^I_m(\xi_2 R)=0$, with $d_1$
and $d_2$ the coefficients to be determined.  This leads to the following
equation for the spin-up case
\begin{equation}\label{det}
    {\xi_1(\xi_2^2+X_1) \over \xi_2(\xi_1^2+X_1)} =
    {I_m(\xi_2 R)I_{m+1}(\xi_1 R) \over I_m(\xi_1 R)I_{m+1}(\xi_2 R)} ,
\end{equation}
from which the eigen state energy can be obtained.
The nature of function $I_m(\xi r)$ guarantees that the amplitude of the
edge state wave function will be large near the disk edge and small in
the disk center.

In the large disk radius limit, i.e. $R\rightarrow \infty$, $I_m(\xi r)$
approaches to $e^{\xi r}/\sqrt{2\pi \xi r}$, a value independent on $m$.
From Eq.(\ref{det}) we obtain eigen state energy $E_0=-DM/B$ for both
spin-up and spin-down states.
For large radius $\xi R>>1$, the eigen state energy is given by
$E_{m,s}=E_0+\delta E_{m,s}$ ($s$ denotes the spin index) with
\begin{equation}\label{limt}
    \delta E_{m,s} \simeq C'_s (m+{1\over 2}) R^{-1} ,
\end{equation}
where $C'_s$ is a factor only relies on the model parameters and spin.
Eq.(\ref{limt}) suggests that for a large radius of the disk, the energy
of the edge states approach linearly in $R^{-1}$ to the limiting value $E_0$,
and for a fixed $R$ one get equal energy spacing for eigen states with
adjacent $m$ index.

\begin{figure}[ht]
\includegraphics[width=6cm,height=9cm,angle=-90]{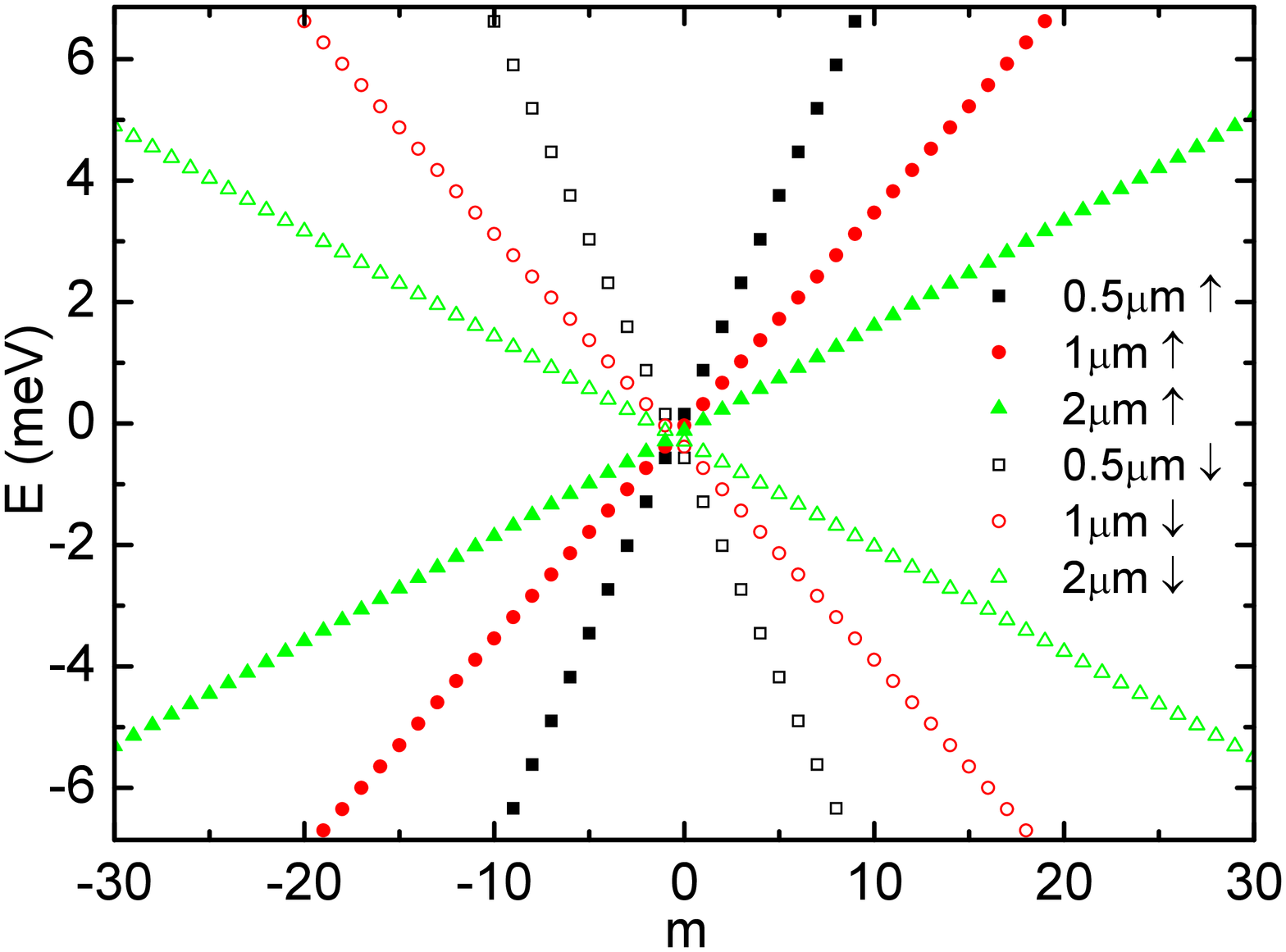}
\caption{(Color online)
The energy spectrum versus the angular momentum index $m$ for
three values of the disk radius $R$: $0.5$ $\mu$m (square dots), $1$ $\mu$m
(round dots), and $2$ $\mu$m (triangular dots).  The energy of spin-up (spin-down)
state is depicted by solid (open) symbols, respectively.
Parameters $A$, $B$, $D$, $M$ are taken from Ref.\cite{Zhang06}:
$A=-342$ meV nm, $B=-169$ meV nm$^2$, $D=5.14$ meV nm$^2$, $M=-6.86$ meV.
}\label{fgengm}
\end{figure}

In general, the eigen state energy denoted as $E_{m,s}(R)$ is a function
of $m$ and $R$.  In Fig.\ref{fgengm}, $E_{m,s}(R)$ is shown as a function
of the angular momentum index $m$ for three values of $R$.
The energy of spin-up states is depicted by solid symbols and the energy
of spin-down states is shown by open symbols.  The energy exhibits an approximately
linear dependence on $m$ and the slope of the curves becomes smaller as the disk radius $R$
increases.  It is found that for each $m$, there is a critical value $R_m^c \ne 0$
such that for $R<R_m^c$ no edge states can exist inside the bulk gap as one can
not find any energy $|E|<|M|$ that Eq.(\ref{det}) holds.  It is also found that $R_m^c$
becomes larger as $m$ increases.
\begin{figure}[ht]
\includegraphics[width=6cm,height=9cm,angle=-90]{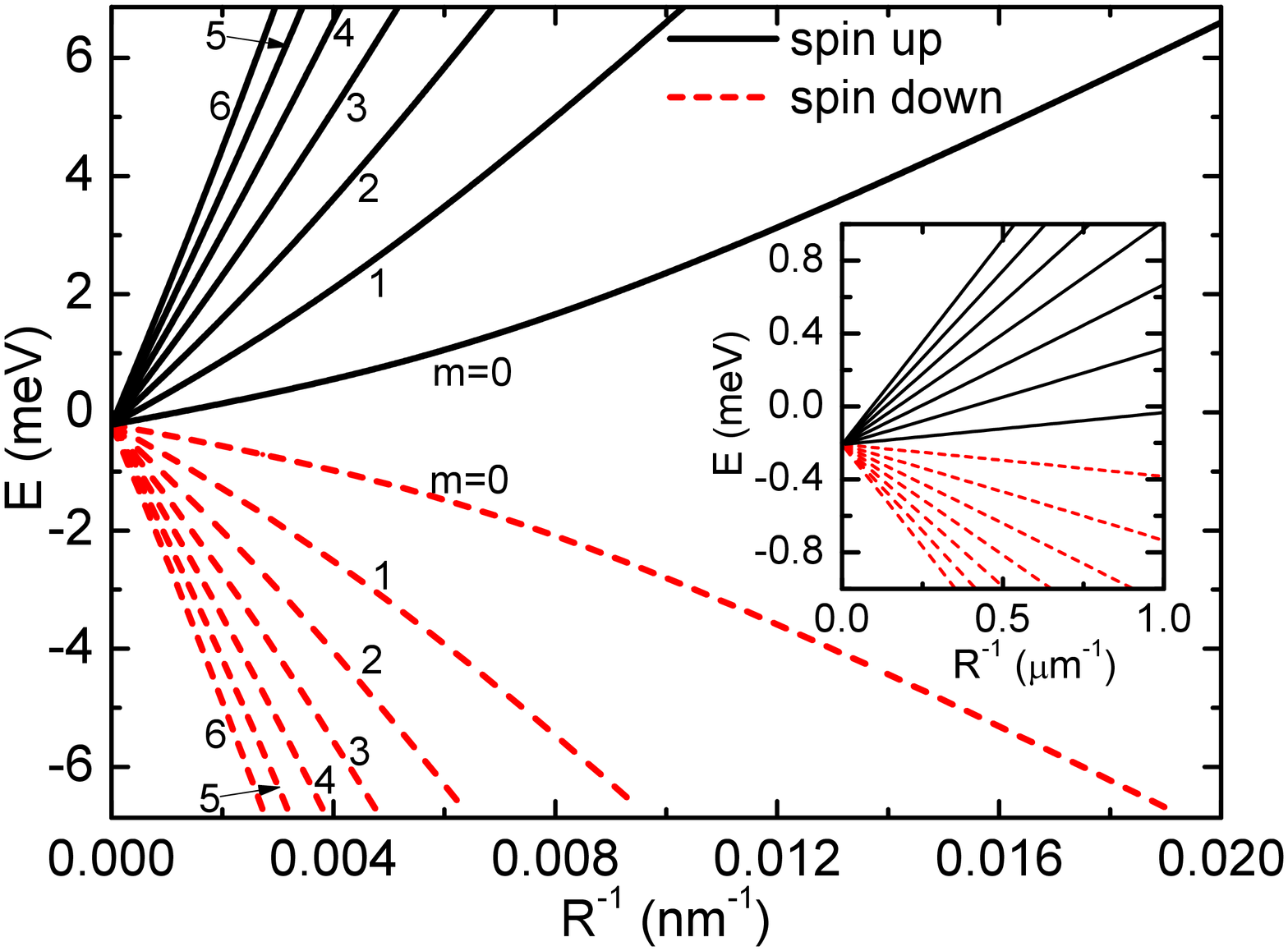}
\caption{(Color online)
The $R^{-1}$ dependence of the edge state energy for several values of $m$.
The black solid curves are for the spin-up states and the red dashed curves
are for the spin-down states.  The inset shows an enlarged portion of the
figure near $R^{-1}\approx 0$.
}\label{fgengr}
\end{figure}

In Fig.\ref{fgengr}, the eigen state energy is shown as a function of $R^{-1}$
for several values of $m$.  It is clear that the $R^{-1}$ dependence is not
linear.  For a fixed $m$, this non-linear dependence is different for spin-up
and spin-down states.

\begin{figure}[ht]
\includegraphics[width=6cm,height=9cm,angle=-90]{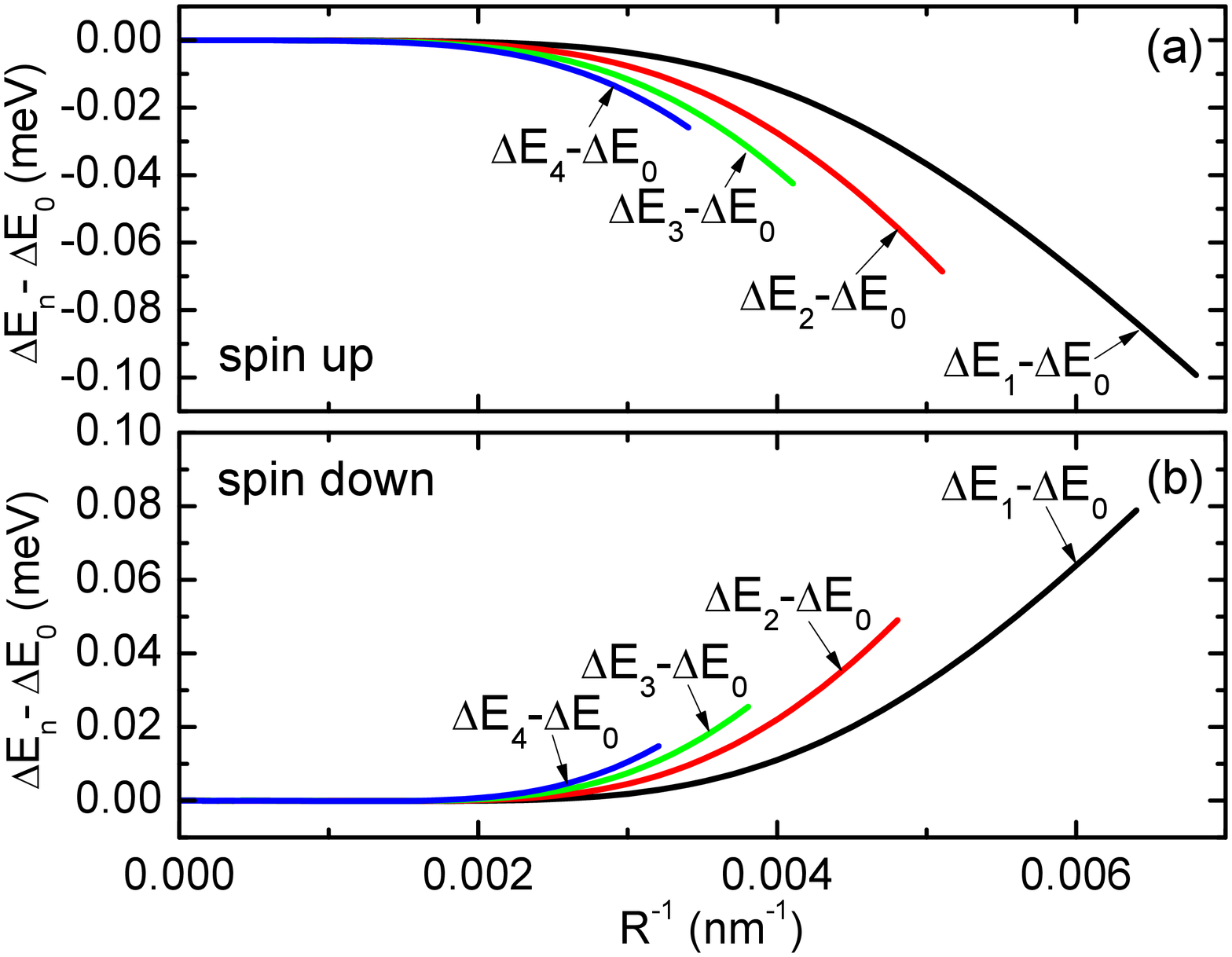}
\caption{(Color online)
The $R^{-1}$ dependence of the energy spacing $\Delta E_{m,s}-\Delta E_{0,s}$.
The upper panel (a) is for the spin-up state and the lower panel (b) is for the
spin-down state.
}\label{fgspacing}
\end{figure}

Eq.(\ref{limt}) shows that for large disk radius the energy spacing $\Delta E_{m,s}
=E_{m+1,s}-E_{m,s}$ depends linearly on $R^{-1}$ and is independent of $m$.
In Fig.\ref{fgspacing}, $\Delta E_{m,s}-\Delta E_{0,s}$ is shown as a function
of $R^{-1}$ for several values of $m$.  The upper panel (a) is for the spin-up
states and the lower panel (b) is for the spin-down states.  The energy spacing
becomes smaller as $m$ increases.  This clearly demonstrates that the $m$
dependence of eigen state energy shown in Fig.1 is non-linear intrinsically.
%

%
%

In the remaining part of this paper, base upon our exact wave functions,
we discuss some interesting aspects of 2DTI systems with circular geometry.
In the case of the disk shaped 2DTI, with the open boundary condition, one may
also seek eigen state with energy outside the bulk gap.  In this case, the
wave function is a linear combination of $\Phi^J_m(\lambda_1 r)$ and $\Phi^I_m(\xi_1 r)$.
Since $J_m(\lambda r)$ is an oscillatory function, but $I_m(\xi r)$ grows exponentially,
the boundary condition results in a larger weight for the $\Phi^J_m(\lambda_1 r)$ term.
Thus, the amplitude of wave function can not be mainly concentrated near the edge and
decay exponentially toward the center of disk.

Let us consider the anti-disk system: a hole in the 2DTI plane.
The potential is given by $V(r)=0$ when $r>R$.
The open boundary condition is adopted.
For energy inside the bulk gap, The wave function is a linear combination of
$\Phi^K_m(\xi_1 r)$ and $\Phi^K_m(\xi_2 r)$.  The wave function will mainly
concentrated near the edge, thus one has an edge state.  When the energy is
outside the bulk gap, the physically allowed wave function
is a linear combinations of $\Phi^K_m(\xi_1 r)$, $\Phi^J_m(\lambda_1 r)$
and $\Phi^Y_m(\lambda_1 r)$.  The boundary condition can only provide two
equations, but there are three coefficients to be determined.  This means that
one can find an eigen state for any energy outside the bulk gap, completely
different from the disk case.  It is also found that for each $m$, there is a
critical value $R_m^c \ne 0$ (different from the disk case) such that for $R<R_m^c$
no edge states can exist inside the bulk gap.  It is found that $R_m^c$
becomes larger as $m$ increases.  This indicates that, an infinite 2DTI may
have no edge state though it has a finite length edge.

Next, we consider a ring-like geometry of the 2DTI, i.e. $V(r)=0$ for $R_1<r<R_2$,
with the open boundary condition adopted for both edges.
When the energy falls inside the bulk gap, wave functions are a linear
combination of $\Phi^K_m(\xi_1 r)$, $\Phi^K_m(\xi_2 r)$, $\Phi^I_m(\xi_1 r)$,
and $\Phi^I_m(\xi_2 r)$.  The boundary conditions give four equations
for the four coefficients to be determined.  One should have a discrete
energy spectrum.  This system is similar to the ribbon structure \cite{Niu08}
but the two edges are different.

Let us finally consider the case that the potential $V(r)$ is a step
function, i.e., $V(r)=V_0\ne 0$ for $r<R$, and $V(r)=0$ otherwise.
This system can be implemented experimentally by depositing metal gates
on top of the 2DTI.  As the potential $V$ takes different values
for $r<R$ and $r>R$, the values of $\lambda$ in Eq.(\ref{lam})
or $\xi$ in Eq.(\ref{xi}) will be different in the different $r$ regions.
In the following, we introduce a superscript $<$ for $\lambda$ and $\xi$
in the region $r<R$, and a superscript $>$ in the region $r>R$.

In the region $r<R$, the allowed contributions to the wave functions are:
(1) $\Phi^I_m(\xi_1^< r)$ and $\Phi^I_m(\xi_2^< r)$, when $|E-V_0|<|M|$
(both $\lambda_1^<$ and $\lambda_2^<$ imaginary); (2) $\Phi^I_m(\xi_1^< r)$ and
$\Phi^J_m(\lambda_1^< r)$ when $|E-V_0|>|M|$ ($\lambda_1^<$ real,
$\lambda_2^<$ imaginary).  For $r>R$, the allowed contributions are:
(1) $\Phi^K_m(\xi_1^> r)$ and $\Phi^K_m(\xi_2^> r)$ when $|E|<|M|$
(both $\lambda_1^>$ and $\lambda_2^>$ imaginary); (2) $\Phi^K_m(\xi_1^> r)$,
$\Phi^J_m(\lambda_1^> r)$ and $\Phi^Y_m(\lambda_1^> r)$ when $|E|>|M|$
($\lambda_1^>$ real, $\lambda_2^>$ imaginary).
The boundary condition is that the wave function and its first order
derivative should be both continuous at $r=R$.  The boundary condition now
leads to four equations.

When $E>|M|$ or $E<-|M|$, the system should have a continuous energy
spectrum, for any $V_0\ne 0$.  This is because that the number of coefficients
needed to be determined is larger than the number of equations due to the boundary
condition.  When $|E|<|M|$, the quantization of energy level is expected.

%
%

In summary, the exact wave functions for the eigen state of a disk-shaped two
dimensional topological insulator is reported.  The property of edge state is
studied.  For a fixed angular momentum index $m$, there is a critical disk radius
below which the edge state is not possible.  The value of this critical radius
increases as $m$ increases.  The $R^{-1}$ dependence of the edge state energy is
non-linear.  The $m$ dependence of the edge state energy is non-linear as well.
The exact wave functions also make it easy for us to investigate the electronic
state in other structures of the two dimensional topological insulator.

%
%

\begin{acknowledgments}

This work was partly supported by NSF and MOST of China.

\end{acknowledgments}

%
%

%
%

\end{document}